# Two-dimensional Intrinsic Janus Structures: Design Principle and Anomalous Nonlinear Optics

Yang Li[1,2], Chengzhi Wu[2], Xuelian Sun[2], Liangting Ye[2], Yirui Lu[2], Hai-Qing Lin[3,2], Wenhui Duan[4,5,6], and Bing Huang[2,7,*]

[1]*School of Mathematics and Physics, University of Science and Technology Beijing, Beijing 100083, China*

[2]*Beijing Computational Science Research Center, Beijing 100193, China*

[3]*Institute for Advanced Study in Physics and School of Physics, Zhejiang University, Hangzhou 310058, China*

[4]*State Key Laboratory of Low Dimensional Quantum Physics, Department of Physics, Tsinghua University, Beijing 100084, China*

[5]*Institute for Advanced Study, Tsinghua University, Beijing 100084, China*

[6]*Frontier Science Center for Quantum Information, Beijing 100084, China*

[7]*School of Physics and Astronomy, Beijing Normal University, Beijing 100875, China*

*Correspondence: bing.huang@csrc.ac.cn

## ABSTRACT

Two-dimensional Janus structures have garnered rapidly growing attention across multidisciplinary fields. However, despite extensive theoretical and experimental efforts, a principle for designing intrinsic Janus materials remains elusive. Here, we propose a first-principles alloy theory based on cluster expansion, incorporating a strong repulsive interaction of a cation-mediated anion-pair cluster and refined short-range cluster-cluster competitions, to unravel the formation mechanism of intrinsic Janus structures with a distorted 1T phase among numerous competing phases. Our theory not only explains why intrinsic Janus structures are accidentally observed in RhSeCl and BiTeI which are composed of alloyed elements from different groups, but also accurately predicts a wide range of 1T-like intrinsic Janus materials that are ready for synthesis. Intriguingly, as demonstrated in the case of RhSeCl, we reveal that intrinsic Janus materials can exhibit anomalous second-harmonic generation (SHG) with a distinct quantum geometric effect, originating from strong lattice and chemical-potential mirror asymmetry. Furthermore, a novel skin effect unexpectedly emerges in finite-thickness RhSeCl, accompanied by a hidden SHG effect within the bulk region. Our theory paves the way for the *ab initio* design of intrinsic Janus materials, significantly accelerating progress in Janus science.

# INTRODUCTION

Designing multifunctional two-dimensional (2D) materials is crucial for exploring fascinating physical phenomena and enhancing material diversity [1-8]. Different from the disordered systems created through chemical functionalization [9, 10] or composite structures formed by van der Waals (vdW) heterostructures [11-14], Janus engineering provides a unique way to effectively design noncentrosymmetric crystalline structures, facilitating the investigation of novel physical phenomena inherent in inversion-asymmetric systems [15-22]. These properties include piezoelectronics [16, 17, 23, 24], Rashba spintronics [25-27], valleytronics [28-30], and multiple topologies [31-33], enabling a wide range of applications in physics, chemistry, energy, and information sciences [34-37]. However, realizing these exciting applications depends on the availability of stable Janus structures, which remains one of the most significant challenges in Janus science.

The 2D Janus structures involve replacing one or more atomic layers with elements from the same or different groups [16-33]. The conventional isovalent substitution provides a rational design strategy. However, a frequently overlooked fact is that the chemical similarity of same-group elements (SGEs) paradoxically promotes disordered alloy formation. Consequently, the SGE Janus phases are typically highly metastable, making them extremely challenging to synthesize in state-of-the-art experiments. For instance, although numerous 2D metastable SGE-Janus materials have been proposed over the past decade, only two (MoSSe [16-18] and PtSSe [20]) have been successfully synthesized through technically demanding processes. On the other hand, it was accidentally discovered that two different-group-element (DGE) Janus phases, RhSeCl [38-40] and BiTeI [21, 25, 41-47], can exist in their ground states. Unlike SGE-Janus materials, these DGE phases are intrinsically stable and, therefore, significantly easier to synthesize on a large scale and even in multiple layers. However, it has long been puzzling why intrinsic Janus materials have only been discovered in these two DGE systems. Furthermore, it remains largely unresolved whether a principle can be established to enable the *ab initio* design of intrinsic Janus structures on demand.

In this article, we propose a first-principles alloy theory based on cluster expansion (CE) to elucidate the unusual formation mechanism of intrinsic Janus materials with 1T-type structure, as demonstrated in the examples of Rh$XY$ family ($X$ = S, Se, Te; $Y$ = Cl, Br, I). Interestingly, we discover that the emergence of a cation-mediated anion-pair cluster with strong repulsive interaction, accompanying with precise short-range cluster-cluster competitions, plays a pivotal role in the formation of intrinsic Janus structures in the 1T-phase DGE systems. In contrast, the absence of this unique anion-pair interaction in all SGE systems ultimately leads to highly metastable Janus structures. Leveraging this theory, we predict 26 novel intrinsic Janus systems that are ready for synthesis. Importantly, due to strong lattice and chemical-potential mirror asymmetry, we reveal that the Rh*XY* family exhibits the anomalous second-harmonic generation (SHG) with a distinct quantum geometric effect, offering a unique platform to manipulate band geometry through nonlinear optical (NLO) methods. Furthermore, a novel skin effect emerges in finite-thickness RhSeCl, accompanied by a hidden SHG effect within the bulk region.

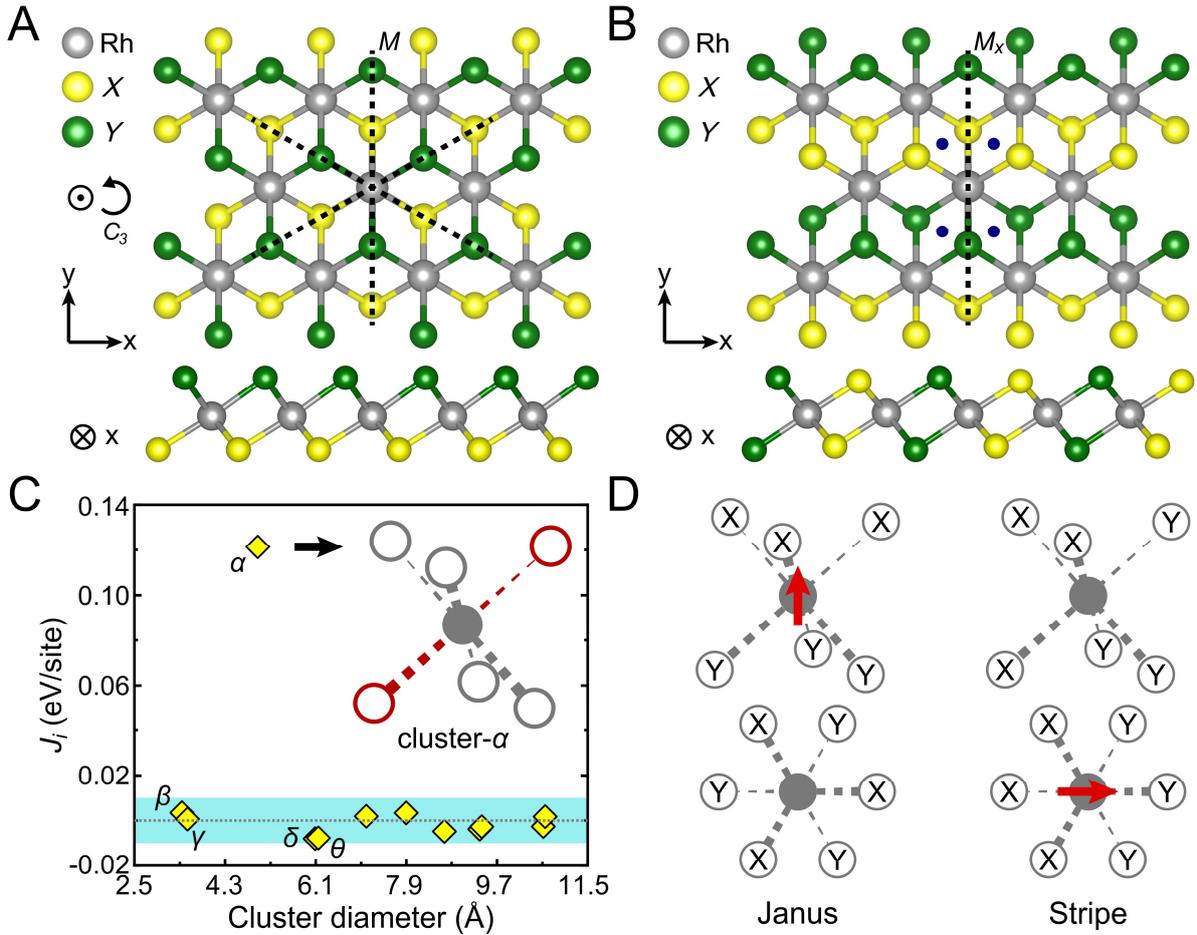

**Figure 1. Formation mechanism of Janus and stripe structures in 1T-phase DGE materials.** (A) and (B) Crystal structures of monolayer Rh*XY* in the Janus and stripe phases, respectively. The upper and lower panels show the top and side views of the crystal structures, respectively. (C) Cluster-expansion coefficient $J_i$ of monolayer RhSeCl as a function of cluster diameter. The inset shows the configuration of dominant pair cluster-$\alpha$ (highlighted in red). (D) Schematic diagrams illustrating the formation of Janus (*Left*) and stripe (*Right*) phases. Red arrows indicate the shift of Rh atoms, leading to a distorted 1T structure.

## RESULTS

**Cluster-expansion alloy theory.** We begin our discussion with monolayer Rh*XY*. To date, only two ordered phases have been experimentally discovered [38-40, 48-50]: the Janus phase [Fig. 1A] and the stripe phase [Fig. 1B]. The Janus phase crystallizes into a triple-layered structure with the *P3m1* space group, where the Rh cationic layer is sandwiched between *X* and *Y* anionic layers. As shown in Fig. 1A, each atomic layer forms a standard triangular lattice, giving the system a three-fold rotational symmetry and three distinct mirror symmetries. The stripe phase, another common phase of Rh*XY* [48-50], belongs to the *P2$_1$/m* space group, where *X* and *Y* atoms are mixed in both upper and lower anionic layers, forming a stripe-like structure. As illustrated in Fig. 1B, the stripe structure has only one mirror operation with respect to the *y*-axis. Unlike Janus phase, it contains four inversion centers within its unit cell, preserving a global inversion symmetry.

The conventional approach of analyzing ordered structures within a given system is insufficient to capture the critical factors governing crystal formation. We recognize that the Janus or stripe Rh*XY* phases can be viewed as distorted 1T-phase alloy structures, where Rh is situated in the octahedral cage formed by *X* and *Y* anions, and the varying occupancies of *X*/*Y* result in different binary alloys. This insight motivates us to employ CE method to explore the most energetically favorable configurations and characteristics of dominant clusters, thereby elucidating the formation mechanism of Janus Rh*XY* among numerous competing structures. After an extensive CE structure search (> 5000 alloy structures for each Rh*XY*), we find that only the Janus and stripe phases can serve as ground states. Specifically, RhSeCl and RhSCl exhibit the Janus phase, while other Rh*XY* systems crystallize into the stripe phase (Section I in Supplementary Information). The CE results not only align with the intrinsic Janus phase of RhSeCl but also match the stripe structure of RhTeCl observed in existing experiments [38-40, 48-50]. Similar results are also observed in BiTeI-family [21, 25] (Section I in Supplementary Information).

The effective cluster interactions (ECIs) determine the contribution of each cluster to the formation energy ($E_f$) of an alloy system. Taking RhSeCl as an example, the converged CE Hamiltonian can be well expressed by only single-site and pair-cluster interactions. In Fig. 1C, the value of ECI ($J_i$) is plotted as a function of cluster diameter. Different from the typical case where $J_i$ decreases with increasing cluster diameter, an exceptionally large positive $J_i$ is observed at the diameter of ~4.95 Å, corresponding to the cluster-*α* composed of two double-sided Rh-mediated anions at the third-nearest-neighboring sites [inset in Fig. 1C]. Crucially, this key pair-cluster is exclusively observed in DGE alloy phases and absent in SGE ones (e.g., MoSSe, WSSe, PtSSe and Cr$_2$Br$_3$I$_3$) (Sections II and III in Supplementary Information), indicating its unusual role in forming intrinsic Janus structures. In RhSeCl, because the two anions belong to different groups, the Rh-Se and Rh-Cl bonds exhibit significant differences, reflecting a general feature of DGE-Janus structures. Notably, the same occupancies of cluster-α will result in a large positive contribution to the CE Hamiltonian, sharply increasing $E_f$ and inevitably causing instability. Consequently, the two sites in cluster-α must be occupied by different anions to minimize the total energy of the system, thereby stabilizing the structure.

Interestingly, the presence of cluster-α leads to only two ordered alloy phases. First, when $X$ atoms occupy all three sites in top anion-layer, the large positive $J_i$ of cluster-α forces $Y$ atoms to occupy the counter-sites in bottom anion-layer, precisely resulting in a DGE-Janus structure [*Left* in Fig. 1D]. Second, when $X$ atoms occupy any two sites in top anion-layer and $Y$ atom occupies the remaining site, only one $Y$ and two $X$ atoms will occupy the three counter-sites in bottom anion-layer to minimize $E_f$, leading to a stripe structure [*Right* in Fig. 1D]. We emphasize that such a mechanism is general for 1T-phase DGE materials, including BiTeI-family (Section I in Supplementary Information). In contrast, the absence of such a cluster-α ultimately results in disordered (random) alloy phases in SGE materials, including 2H-MoSSe, 2H-WSSe, 1T-PtSSe and magnetic $Cr_2Br_3I_3$. (Sections II and III in Supplementary Information).

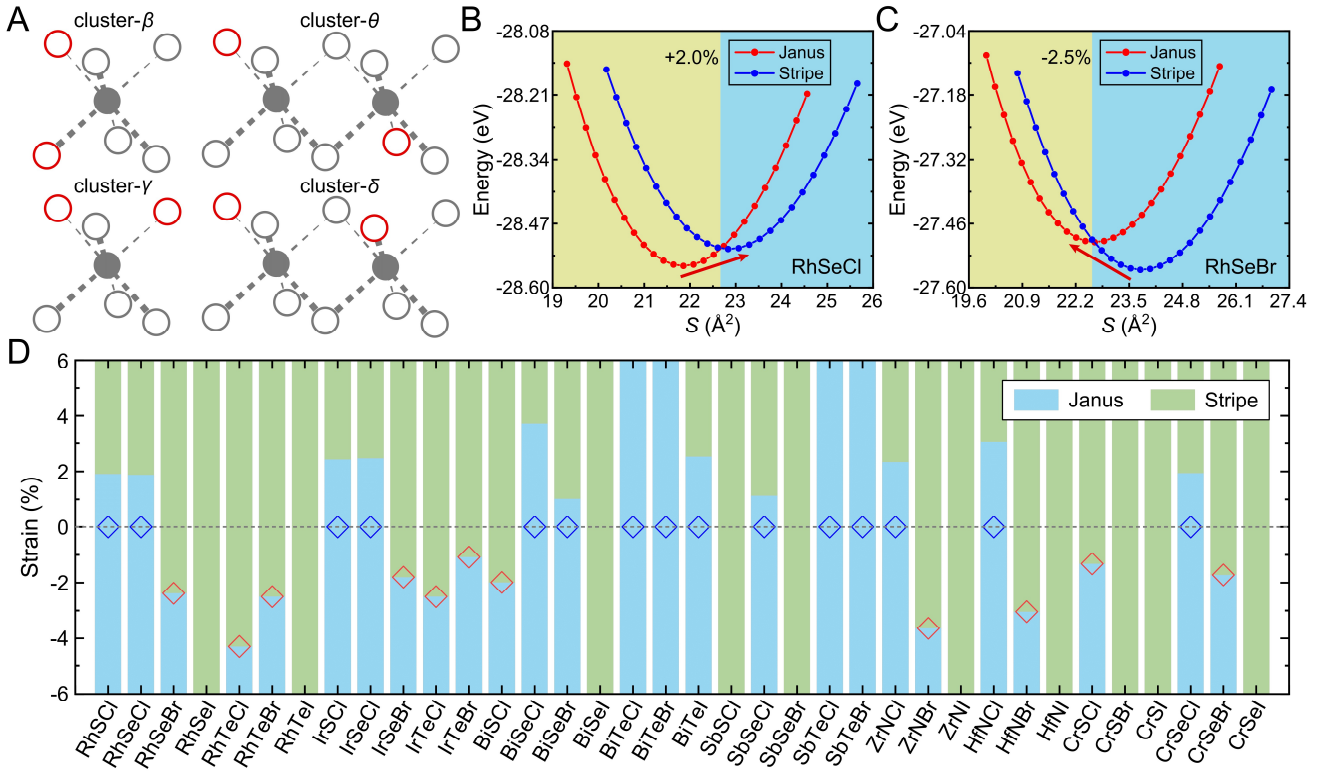

**Figure 2. Structural phase transition between Janus and stripe structures and prediction of intrinsic Janus materials.** (A) Configurations of four major short-range pair-clusters marked in Fig. 1C. (B) and (C) Total energies as a function of effective area $S$ for monolayer RhSeCl and RhSeBr in both Janus and stripe structures, respectively.

Red arrows indicate the direction of strain-induced phase transition. (D) Formation range of Janus and stripe structures for 36 1T-type materials under external strain.

Contrary to the cluster-$\alpha$, the $J_i$ of other clusters are an order of magnitude weaker. Figure 2A shows the configurations of other four primary short-range pair-clusters [marked as $\beta$, $\gamma$, $\theta$, and $\delta$ in Fig. 1C]. Interestingly, the $J_i$ of these clusters have opposite signs. The cluster-$\beta$ and cluster-$\theta$ are consist of two sites in different anion-layers, while the cluster-$\gamma$ and cluster-$\delta$ are formed by two sites in the same anion-layer. These short-range pair-clusters are universal in all the 1T-phase alloys. Due to their small $J_i$ values, they have a very weak impact on the absolute $E_f$ of Rh$XY$. However, the competition between these small, opposite $J_i$ can precisely determine the ground state as Janus or stripe structure, as these two phases exhibit only a small $E_f$ difference. In practice, the values of ECIs may be manipulated by an external field. Figure 2B shows the $E_f$ curvatures of RhSeCl in both Janus and stripe structures as a function of strain, where the strain values are converted to effective area ($S$) for comparison. Interestingly, there is a crossover of the energy curves at $S \approx 22.6$ Å ($\varepsilon \sim 2.0\%$), i.e., when $\varepsilon > 2.0\%$, the ground-state can be converted from Janus phase to stripe phase in RhSeCl. Conversely, as shown in Fig. 2C, the ground-state of RhSeBr can be converted from stripe phase to Janus phase when $\varepsilon < -2.5\%$. This intriguing discovery suggests that the ground-state of Rh$XY$ can be designed on demand by using different epitaxial substrates, owing to strain-controlled $J_i$ of $\beta$, $\gamma$, $\theta$, and $\delta$ clusters.

The above understanding can be summarized into a unified theory for the formation of intrinsic Janus structures in 1T-phase lattice: *the emergence of the unusual cluster-α in DGE alloys drives systems into two possible ground-states (Janus or stripe phases), the precise competitions of β, γ, θ, and δ clusters ultimately selects one of these two*. This theory can be further applied to significantly expand the library of intrinsic Janus materials. For example, Fig. 2D shows the formation range of Janus and stripe structures for 36 1T-type materials, which cover the typical 2D DGE materials, including (Rh/Ir)SeCl-family [38-40, 48-50], (Bi/Sb)TeI-family [21, 25, 41-47], transition-metal nitride halides [51], and ferromagnetic CrSCl-family [52-56] (See section IV in Supplementary Information for more discussions). Among these materials, 15 candidates are intrinsic Janus ones, and another 11 candidates can transform into Janus structures under appropriate strain. It is worth noting that Janus materials BiTeCl, BiTeBr, and CrSCl have been synthesized recently [21, 56], providing experimental support for the broad applicability of this unified theory.

It should be remarked that the unique local structure of 1T-phase lattice is crucial for cluster-expansion alloy theory. Within this coordination environment, the cation forms nearly 180° bond angles with its two chemically bonded anions, giving rise to strong $\sigma$ bonds within the cation-mediated anion pair, i.e., cluster-$\alpha$. This renders cluster-$\alpha$ an exceptionally stable building block, which subsequently drives the formation of Janus or stripe phases. In contrast, such a local structure is absent in other typical 2D structures, e.g., 2H phase and $MX_3$ phase, which may explain why intrinsic Janus structures have so far been observed exclusively in the materials with 1T-type phase.

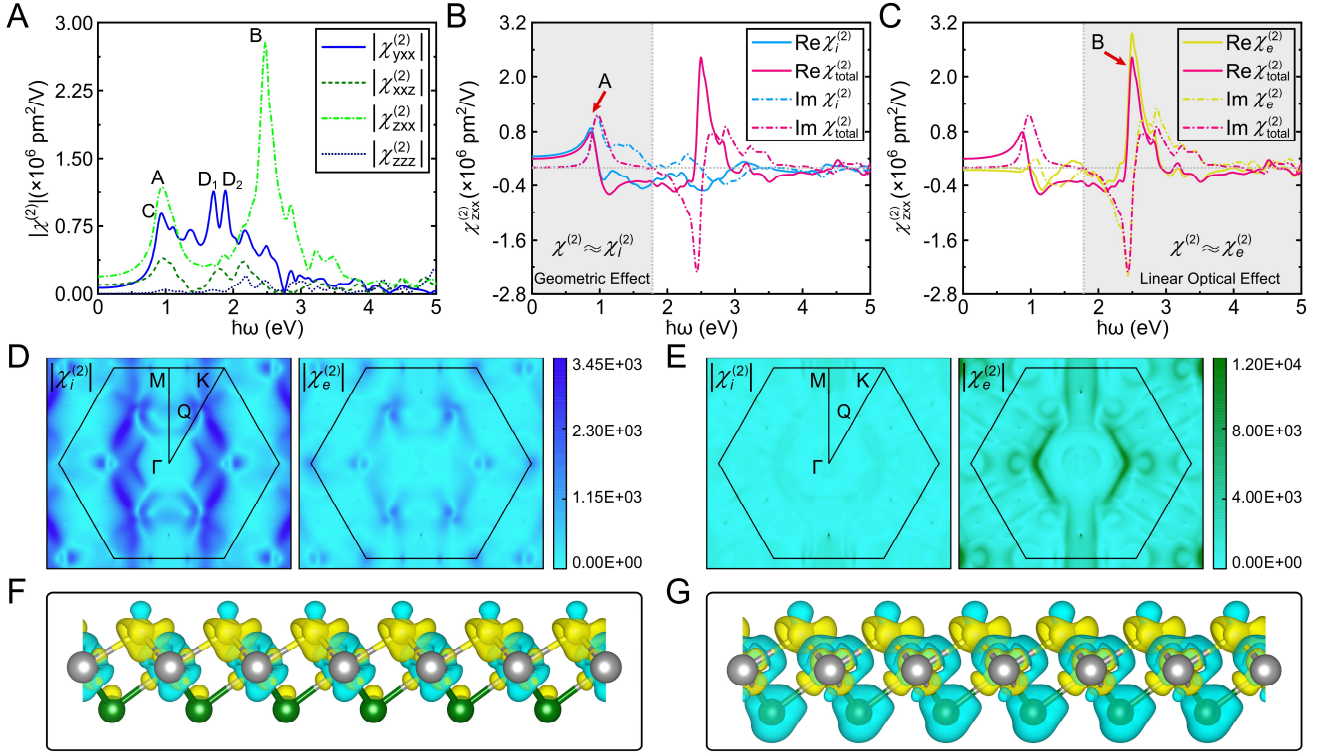

**Figure 3. Anomalous SHG response of monolayer RhSeCl.** (A) Four independent components of the second-order nonlinear susceptibility, $\chi^{(2)}$, as a function of incident photon energy for monolayer RhSeCl. (B) and (C) Contributions of frequency-dependent $\chi_i^{(2)}$ and $\chi_e^{(2)}$ to $\chi_{zxx}^{(2)}$ for RhSeCl, respectively. Here, gray-vertical-dash line distinguishes the frequency regions of $\chi_{zxx}^{(2)}$ caused by different mechanisms. (D) and (E) K-resolved $\chi_i^{(2)}$ (*Left*) and $\chi_e^{(2)}$ (*Right*) of the peaks A and B of $\chi_{zxx}^{(2)}$, respectively. Here, black-hexagon shows the Brillouin Zone. (F) Differential charge density between $VB_1$ and $CB_1$. (G) Summation of differential charge density between $VB_4$ and $CB_2$ and differential charge density between $VB_5$ and $CB_2$. Here, $VB_n$ ($CB_n$) indicates the *n*th valence (conduction) band from band edges, and the iso-surface values are set to 0.0068 to clearly display the charge density distribution.

**Anomalous SHG response.** It is important to further explore the unique physics exclusive to intrinsic Janus materials. We note that the key feature of DGE-Janus structure is the combination of strong lattice and chemical-potential mirror asymmetry, which could generate anomalous NLO responses (e.g., SHG) barely seen in other systems. When light with frequency of $\omega$ passes through a material, its frequency will be doubled via SHG process with the strength of $\chi^{(2)}_{abc}$. Taking monolayer RhSeCl as an example, the *P3m1* lattice symmetry results in four independent nonzero susceptibility components: $\chi^{(2)}_{xxy} = \chi^{(2)}_{yxx} = -\chi^{(2)}_{yyy}$, $\chi^{(2)}_{zxx} = \chi^{(2)}_{zyy}$, $\chi^{(2)}_{xxz} = \chi^{(2)}_{yyz}$, and $\chi^{(2)}_{zzz}$. As shown in Fig. 3A, the $\chi^{(2)}_{xxz}$ and $\chi^{(2)}_{zzz}$ are relatively small over the entire energy range. For RhSeCl, the in-plane component $\chi^{(2)}_{yxx}$ shows significant value within a wide range of 0.0 eV < $\hbar\omega$ < 3.0 eV, with noticeable peaks of ~0.88×10$^6$ pm$^2$/V at $\hbar\omega$ = 0.9 eV (peak C) and ~1.14×10$^6$ pm$^2$/V around $\hbar\omega$ = 1.8 eV (peaks D$_1$ and D$_2$). Meanwhile, the out-of-plane component $\chi^{(2)}_{zxx}$ has only two major peaks (peaks A and B), with the values of ~1.18×10$^6$ pm$^2$/V at $\hbar\omega$ = 0.9 eV and ~2.78×10$^6$ pm$^2$/V at $\hbar\omega$ = 2.5 eV. Remarkably, the SHG peaks in RhSeCl are significant, ~5 times that of MoS$_2$ or MoSSe [57, 58], ~10 times that of NbOCl$_2$ and NbOI$_2$ [59-61], and even ~100 times that of BN [62, 63]. Moreover, the static SHG of RhSeCl is also comparable to those of prominent 2D NLO materials, further demonstrating the outstanding performance for the SHG responses of RhSeCl (Sections V and VI in Supplementary Information). In addition, a similar SHG intensity is also observed in other Janus Rh*XY* (Section VII in Supplementary Information).

The contributions to SHG can be categorized into two distinct terms: the purely interband transition term $\chi^{(2)}_e$ and the mixed intraband and interband transition term $\chi^{(2)}_i$. The $\chi^{(2)}_e$ depends exclusively on transition-dipole moment $r^a_{nm}$ defined as the Berry connection $A^a_{nm} \equiv i\langle u_n|\partial_{k_a} u_m\rangle$, reflecting the linear optical effect. Alternatively, the $\chi^{(2)}_i$ is closely related to the generalized derivative $r^a_{nm;b} = \partial_{k_a} r^a_{nm} - ir^a_{nm}(A^a_{nn} - A^a_{mm})$, the quantum metric $g^{ab}_{nm} = \text{Re}(r^a_{nm} r^b_{mn})$, and the Berry curvature $\Omega^{ab}_{nm} = \text{Im}(r^a_{nm} r^b_{mn})$, which arises from the unusual geometric effects [64]. In conventional 2D systems, e.g., MoS$_2$ and SGE-Janus MoSSe, these two effects are strongly intertwined, hindering the optical manipulation of quantum geometry in 2D limit. Remarkably, we find that in DGE-Janus systems, the quantum geometric effect can be effectively decoupled from the linear optical effect. As for RhSeCl, when 0.0 eV < $\hbar\omega$ < 1.7 eV, $\chi^{(2)}_{zxx}$ is governed by $\chi^{(2)}_i$ [Fig. 3B]; when $\hbar\omega$ > 1.7 eV, $\chi^{(2)}_{zxx}$ is dominated by $\chi^{(2)}_e$ [Fig. 3C]. Additionally, the quantum geometric contributions can be resolved in momentum space during the NLO process. For example, we calculate the K-resolved $\chi^{(2)}_e$ and $\chi^{(2)}_i$ of RhSeCl for peaks A and B. At peak A, $\chi^{(2)}_e$ is significant at most positions in the first Brillouin Zone (BZ), while $\chi^{(2)}_i$ has weak contributions only near the Q point [Fig. 3D]. Conversely, at peak B, $\chi^{(2)}_e$ has little contribution over the entire BZ, while $\chi^{(2)}_i$ has obvious contributions around the Γ point [Fig. 3E]. A similar anomalous phenomenon is also obtained in the in-plane component $\chi^{(2)}_{yxx}$ (Sections V and VIII in Supplementary Information).

To confirm that this anomalous SHG originates from the unique characteristics of the DGE-Janus structure, we further calculate the differential charge density between the optical-excited conduction and valence bands. For RhSeCl, the electronic states contributing to peak A, the major peak in the quantum geometric region, exhibit a strong charge-transfer-imbalance between top-anion-layer and bottom-anion-layer, reflecting the strong mirror asymmetry of the chemical potential in a Janus structure [Fig. 3F]. In contrast, such a novel phenomenon cannot be observed in those electronic states contributing to peak B [Fig. 3G]. Moreover, this anomalous SHG cannot be observed in the SGE-Janus structure (Section IX in Supplementary Information), because it lacks the DGE-induced strong chemical-potential imbalance. Thus, the DGE-Janus structure allows for a significant difference in charge transfer during the optical transition, leading to a distinct quantum geometric region in the NLO process.

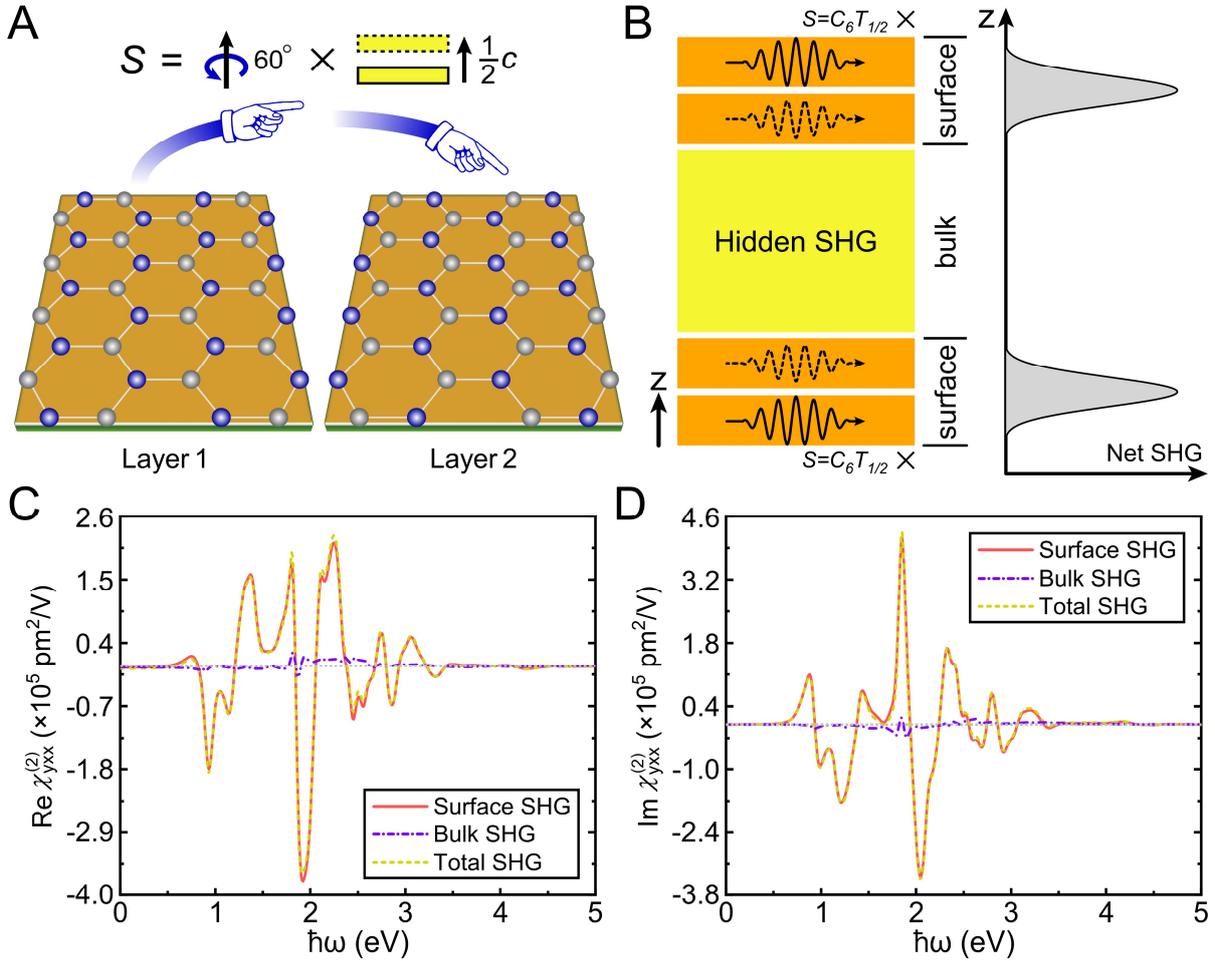

**Figure 4. Novel skin effect in finite-thickness RhSeCl.** (A) Schematic diagram of relationship between two adjacent layers in bulk RhSeCl. Here, blue and gray balls represent Se and Cl atoms, respectively. (B) Schematic diagram of the skin effect of the in-plane component $\chi^{(2)}_{yxx}$ in RhSeCl film. (C) and (D) Comparison of surface and bulk

contributions to the real and imaginary parts of $\chi^{(2)}_{yxx}$ in a 20-layers RhSeCl slab, respectively. Here, we define the five layers near the top and bottom vacuum regions of the slab model as the surface layers.

**Novel skin effect.** In experiments, the bulk RhSeCl crystallizes into a hexagonal structure with the *P6₃mc* space group [38-40]. This crystal symmetry results in three nonvanishing susceptibility components: $\chi^{(2)}_{zxx} = \chi^{(2)}_{zyy}$, $\chi^{(2)}_{xxz} = \chi^{(2)}_{yyz}$, and $\chi^{(2)}_{zzz}$, among which the component $\chi^{(2)}_{zxx}$ attains a value as large as ~3.17×10³ pm/V at $\hbar\omega = 0.9$ eV, indicating the pronounced NLO performance of bulk RhSeCl (Section X in Supplementary Information). Notably, compared with the monolayer case, the key distinction is the absence of the in-plane component $\chi^{(2)}_{yxx}$ in the bulk phase (Section X in Supplementary Information). As illustrated in Fig. 4A, the adjacent layers in bulk RhSeCl are connected via a 60° in-plane rotation. This specific rotational arrangement leads to opposite $\chi^{(2)}_{yxx}$ in adjacent layers, resulting in a net cancellation of the in-plane SHG response. This phenomenon gives rise to a hidden NLO effect in an *inversion-asymmetric* system, which differs fundamentally from the hidden effects observed in *inversion-symmetric* systems (Section XI in Supplementary Information).

The absence of $\chi^{(2)}_{yxx}$ in the bulk phase, coupled with its presence in the monolayer, motivates further investigation into the mechanism of the anomalous surface effect contributing to the SHG response. In bulk RhSeCl, the lack of component $\chi^{(2)}_{yxx}$ is primarily attributed to the $S = C_6 T_{1/2}$ symmetry, where $C_6$ represents a six-fold rotation and $T_{1/2}$ denotes a lattice translation by 1/2 along the out-of-plane direction. At the surface, the $S$ symmetry is broken, enabling the emergence of a nonzero $\chi^{(2)}_{yxx}$. Consequently, the $\chi^{(2)}_{yxx}$, which has no net contribution in regions far from the surface, accumulates near the surface, giving rise to a novel skin effect [Fig. 4B]. To validate this, we calculated the layer-resolved $\chi^{(2)}_{yxx}$ for a 20-layer-thickness RhSeCl slab. As shown in Figs. 4C and 4D, the summation of $\chi^{(2)}_{yxx}$ for the 1-5 layers near the upper surface and 16-20 layers near the lower surface accounts for nearly the entire $\chi^{(2)}_{yxx}$ of RhSeCl slab, while the summation of $\chi^{(2)}_{yxx}$ for 5-15 layers in the middle of the slab is approximately zero. Accordingly, the penetration depth of $\chi^{(2)}_{yxx}$ in RhSeCl film is estimated to be ~4 layers (~20 Å) (Section XII in Supplementary Information). Very recently, a similar skin effect of NLO response was reported in the vdW layered antiferromagnets [65], which is attributed to local inversion-symmetry breaking at the surface. Compared with the skin effect in antiferromagnets, the skin effect in RhSeCl raises from a fundamentally distinct physical mechanism, where the interior of RhSeCl exhibits the hidden SHG instead of zero SHG. It should be emphasized that the skin effect in finite-thickness RhSeCl is fundamentally rooted in its crystal symmetry. Although the exciton effect can quantitatively modulate the SHG intensity, in particular around the band edge, the existence of the skin effect is independent of excitonic contributions. Moreover, this novel skin effect is also expected to manifest broadly in other second-order responses of RhSeCl, such as the bulk photovoltaic effect and nonlinear Hall effect.

## DISCUSSION AND SUMMARY

In general, CE is employed for investigating disordered alloy systems. However, it can provide unique insights when applied to a crystalline system, as presented in our case. Conventional structure searches assess stability solely through total energy comparisons. In contrast, CE reveals fundamental formation mechanisms, bridging the gap between theoretical prediction and experimental synthesis. For instance, it is found that the evolutionary algorithm-based searches identify the Janus phase of RhSeCl as the energy minimum, whereas CE uncovers the driving mechanism, i.e., strong repulsive interaction within cluster-$\alpha$ and refined short-range cluster competitions. More importantly, beyond traditional single-configuration analysis, CE also provides a unified Hamiltonian framework. This enables simultaneous treatment of multiple structurally related phases, e.g., Janus and stripe structures in 1T-phase DGE materials, as distinct lattice occupation configurations. Therefore, CE can elucidate the intrinsic physical connections and formation principles behind these different structures, propelling the fundamental understanding essential for structural phase transition and materials design.

While CE theory demonstrates that the 1T-phase DGE compounds exclusively adopt Janus or stripe ground states, the formation of 1T-like materials remains contingent on specific anion-pair compatibility, i.e., not all hetero-anionic combinations satisfy the structural prerequisites. To enable more efficient intrinsic Janus materials design, integrating the cluster-expansion alloy theory with structural search methods provides a reliable approach. Alternatively, the valence-balanced multi-anion substitution provides another promising route. As an example, PtAsCl, which is computationally derived from 1T-PtSe$_2$, can realize its Janus phase via the application of ~5.0% compressive strain to the stripe ground state. This strategy has recently enabled the prediction of altermagnetic DGE-Janus materials [66], confirming its predictive capability. Based on these design principles, more intrinsic Janus materials may be predicted and synthesized, thereby facilitating study of emergent physical phenomena.

In summary, we propose a first-principles alloy theory to reveal the formation mechanism of intrinsic Janus materials with the distorted 1T structures. Our theory provides a clear explanation for why intrinsic Janus structures are solely observed in DGE materials rather than SGE ones. Using RhSeCl as a case study, we demonstrate that intrinsic Janus materials can exhibit an anomalous SHG response and a novel skin effect at finite thicknesses. These findings establish intrinsic DGE-Janus systems as a unique platform for exploring the interplay between quantum geometry, hidden effects, and skin effects within a single material system using NLO methods. This study paves the way for the *ab initio* design of intrinsic Janus structures, advancing the field of 2D Janus science.

## METHODS

**Density functional theory calculation.** The density functional theory (DFT) calculations in our work were performed using the Vienna ab initio Simulation Package (VASP) [67]. Here, the Perdew-Burke-Ernzerhof (PBE) functional [68] was used to approximate the exchange correlation functional in Kohn-Sham equation [69], and the projector augmented wave (PAW) method was used to treat core electrons [70]. The energy cutoff for the plane wave basis was set to 400 eV to ensure the accuracy of results for all materials considered in this work. The convergence criterion for the force on each atom was set to 0.001 eV/Å, and the convergence criterion for total energy was set to $1.0 \times 10^{-8}$ eV for all calculations. In our calculations, $18 \times 18 \times 1$, $18 \times 10 \times 1$, and $18 \times 18 \times 6$ Γ-centered uniform k-mesh were adopted to perform DFT calculations for Janus structure, stripe structure and 3D bulk materials, respectively. For 2D systems, a > 20 Å vacuum layer was used to avoid the artificial interaction caused by periodic boundary conditions. For multilayer systems and 3D bulk materials, the DFT-D3 method was adopted to correct the interlayer van der Waals (vdW) interaction [71]. For magnetic systems, the rotationally invariant generalized gradient approximation (GGA) + U approach [72] was employed to correct the strong correlation effect raised by the partially occupied 3*d* orbitals.

**Cluster expansion.** The basic idea of cluster expansion (CE) is to express the formation energy ($E_f$) of a fixed lattice to a polynomial in the occupation variables [73, 74]

$$E_f(\{S_q\}) = E_0 + \sum_i m_i J_i \langle \prod_{q \in i} S_q \rangle,$$

where, $S_q = +1$ ($S_q = -1$) represents the site $q$ occupied by *X*-type (*Y*-type) atom. Here, $i = \{S_q\}$ represents a cluster, $m_i$ represents the degeneracy of a cluster, $E_0$ represents the total energy of the fully disordered structure. The summation is taken over all symmetry-inequivalent clusters, and the average of correlation function $\prod_{q \in i} S_q$ is taken over all clusters that are symmetry-equivalent to $i$ cluster. $J_i$ is effective cluster interaction (ECI), which can be obtained by fitting the DFT calculated $E_f$ of several pre-constructed structures. Using the well-fitted $J_i$, it is convenient to predict $E_f$ of any structure formed by two different elements and analyze the contributions of every cluster to $E_f$, thus revealing the origin and properties of ground state.

The CE were performed using the Alloy Theoretic Automated Toolkit (ATAT) package [75], with a fixed concentration ratio of 1:1 for all monolayer alloys. The well converged CE for different materials were obtained by fitting the calculated $E_f$ of 36-44 symmetrically inequivalent structures, and the resulting models exhibit cross-validation scores of 0.0003-0.0116 eV/atom, demonstrating their excellent capability in describing these alloy systems. In our calculations, the density of k-point mesh was set to 1800 per reciprocal atom, and the convergence criterion of the total energy was set to $1.0 \times 10^{-6}$ eV.

**Nonlinear optical properties calculations.** The second-harmonic generation (SHG) induced by an incident light $E$ with the frequency of $\omega$ can be described as [76-78]

$$P^a(2\omega) = \varepsilon_0 \chi^{(2)}_{abc} E^b(\omega) E^c(\omega),$$

where $P$ is second-harmonic dipole, $\varepsilon_0$ is vacuum permittivity. Three-order tensor $\chi^{(2)}_{abc}$ is known as the second-order susceptibility with the indices $a$, $b$, $c$ in Cartesian coordinates, which describes the strength of SHG response. In general, $\chi^{(2)}_{abc}$ can be further divided into the purely interband transition term and the mixed intraband and interband transition term, where the purely interband transition term $\chi^{(2)}_{e;abc}$ can be expressed as

$$\chi^{(2)}_{e;abc}(-2\omega, \omega, \omega) = \frac{e^3}{\hbar^2 \Omega} \sum_{nml,k} \frac{r^a_{nm}\{r^b_{ml} r^c_{ln}\}}{(\omega_{ln} - \omega_{ml})} \left[ \frac{2f_{nm}}{\omega_{mn} - 2\omega} + \frac{f_{ln}}{\omega_{ln} - \omega} + \frac{f_{ml}}{\omega_{ml} - \omega} \right],$$

and the mixed intraband and interband transition term $\chi^{(2)}_{i;abc}$ can be written as

$$\chi^{(2)}_{i;abc}(-2\omega, \omega, \omega)$$
$$= \frac{i}{2} \frac{e^3}{\hbar^2 \Omega} \sum_{nm,k} f_{nm} \left[ \frac{2 r^a_{nm}(r^b_{mn;c} + r^c_{mn;b})}{\omega_{mn}(\omega_{mn} - 2\omega)} + \frac{(r^a_{nm;c} r^b_{mn} + r^a_{nm;b} r^c_{mn})}{\omega_{mn}(\omega_{mn} - \omega)} \right.$$
$$\left. + \frac{r^a_{nm}(r^b_{mn} \Delta^c_{mn} + r^c_{mn} \Delta^b_{mn})}{\omega^2_{mn}} \left( \frac{1}{\omega_{mn} - \omega} - \frac{4}{\omega_{mn} - 2\omega} \right) - \frac{(r^b_{nm;a} r^c_{mn} + r^c_{nm;a} r^b_{mn})}{2\omega_{mn}(\omega_{mn} - \omega)} \right].$$

Here, $f_{mn} = f_m - f_n$, $f_m(f_n)$ is the Fermi-Dirac distribution function; $\omega_{mn} = \omega_m - \omega_n$, $\hbar\omega_m(\hbar\omega_n)$ is the eigenvalue of blöch wavefunction; $\Delta^a_{mn} = v^a_{mm} - v^a_{nn}$, $v^a_{mm}(v^a_{nn})$ is the group velocity of electronic states. The transition-dipole moment $r^a_{nm}$ can be calculated using DFT calculations or tight-binding method, and the generalized derivative $r^a_{nm;b}$ can be calculated using the summation rule

$$r^b_{nm;a} = \frac{r^a_{nm} \Delta^b_{mn} + r^b_{nm} \Delta^a_{mn}}{\omega_{nm}} + \frac{i}{\omega_{nm}} \sum_l (\omega_{lm} r^a_{nl} r^b_{lm} - \omega_{nl} r^b_{nl} r^a_{lm}).$$

In our work, the nonlinear optical (NLO) responses calculations were performed using our homemade package NOPSS [79, 80]. To obtain converged results, the density of k-mesh and the number of bands were carefully tested, and the 54 × 54 × 1 Γ-centered uniform k-mesh and 48 electronic bands were used to perform calculations. The energy range of the incident light was set to 0.0-5.0 eV for all calculations, with 500 energy points linearly interpolated within this range. To avoid the divergence, a small imaginary smearing factor $\eta = 0.05$ was added to the frequency of the incident light, i.e., $\omega \to \omega + i\eta$.

**Tight-binding model and projected nonlinear optical properties.** Tight-binding model of RhSeCl slab was constructed using the projected Wannier functions as implemented in the Wannier90 package [81]. To generate the projected Wannier functions, the calculated Blöch wavefunctions were projected into the Rh-3$d$, Se-4$p$, and Cl-2$p$ atomic orbitals. Based on the effective Hamiltonian matrix, the NLO properties were calculated using the Wannier interpolated scheme [77, 82]. In our work, the 500 × 500 × 1 uniform k-mesh was used to obtain the converged SHG, and small imaginary smearing factor $\eta$ of 0.05 was used to avoid divergence. The layer-projected SHG was calculated by left-multiplying the velocity operator with a layer-projection operator. Here, the layer-projection operator is defined as $P_l = \sum_{i \in l} |\phi_i\rangle\langle\phi_i|$, where $l$ is the layer index, $|\phi_i\rangle$ is the projected Wannier function [83, 84]. This projection approach allows for the clear distinction of SHG associated with each layer and ensures that the summation of SHG from each layer is equal to total SHG.

## ACKNOWLEDGEMENTS
This work is supported by National Natural Science Foundation of China (NSFC) (Grants No. W2511008, 12404260, 12347128, and 12088101), National Key Research and Development Program of China (Grant No. 2022YFA1402401), and Science Challenge Project (Grant No. TZ2025013). We acknowledge Dr. Hanpu Liang for technical support and Drs. Ze-Yu Jiang, Sheng-Nan Xu, and Xiao Jiang for helpful discussions. Parts of the calculations of this work were performed at Tianhe-JK at Beijing Computational Science Research Center (CSRC).

## AUTHOR CONTRIBUTIONS
Y.L. and B.H. designed research; Y.L. and C.W. performed research; Y.L. and C.W. contributed new reagents/analytic tools; Y.L. and B.H. wrote the paper; All authors participated in the data analysis and discussion.

## DECLARATION OF INTERESTS

The authors declare no competing interests.

## SUPPLEMENTAL INFORMATION

Supplemental information includes more discussions about the CE calculations of RhSeCl-family and BiTeI-family materials, stability of Janus MoSSe and Janus PtSSe, fitted effective cluster interactions of MoSSe, WSSe, PtSSe, $Cr_2Br_3I_3$ monolayer alloys and several 1T-phase DGE materials, band structure of RhSeCl, second-harmonic generation of RhSeBr and RhTeCl, physical origins of the second-harmonic generation for $MoS_2$ and Janus MoSSe, comparison of second-harmonic generation between RhSeCl and other typical nonlinear optical materials, classification of nonlinear optical responses, and penetration depth of skin effect in RhSeCl.

Supplemental information: Sections I-XII, Figures S1-S18, and Tables S1-S6.